\documentclass[%
aip,
amsmath,amssymb,
reprint,%
]{revtex4-2}

\usepackage{graphicx}
\usepackage{dcolumn}
\usepackage{bm}

\usepackage[utf8]{inputenc}
\usepackage[T1]{fontenc}
\usepackage{mathptmx}
\usepackage{etoolbox}
\usepackage{booktabs}
\usepackage{color,soul}
\usepackage{hyperref}

\def\eg{\emph{e.g.}}
\def\ie{\emph{i.e.}}

\newcommand*{\citen}[1]{%
	\begingroup
	\romannumeral-`\x 
	\setcitestyle{numbers}%
	\cite{#1}%
	\endgroup   
}

\graphicspath{{figs/}}					

\makeatletter
\def\@email#1#2{%
	\endgroup
	\patchcmd{\titleblock@produce}
	{\frontmatter@RRAPformat}
	{\frontmatter@RRAPformat{\produce@RRAP{*#1\href{mailto:#2}{#2}}}\frontmatter@RRAPformat}
	{}{}
}%
\makeatother
\begin{document}
	
\preprint{AIP/123-QED}

\title[Impact of random and targeted disruptions on information diffusion during outbreaks]{Impact of random and targeted disruptions on information diffusion during outbreaks}
\author{Hosein Masoomy}
\homepage{hoseingmasoomy@gmail.com}
\affiliation{
	Dept.\ of Physics, Shahid Beheshti University, 1983969411, Tehran, Iran
}

\author{Tom Chou}%
\homepage{tomchou@ucla.edu}
\affiliation{ 
	Depts.~of Computational Medicine and Mathematics, UCLA, Los Angeles, CA 90095
}

\author{Lucas Böttcher}
\homepage{l.boettcher@fs.de}
\affiliation{
Centre for Human and Machine Intelligence, Frankfurt School of Finance and Management, 60322 Frankfurt am Main, Germany
}%

\date{\today}

\begin{abstract}
Outbreaks are complex multi-scale processes that are impacted not only
by cellular dynamics and the ability of pathogens to effectively
reproduce and spread, but also by population-level dynamics and the
effectiveness of mitigation measures. A timely exchange of information
related to the spread of novel pathogens, stay-at-home orders, and
other containment measures can be effective at containing an
infectious disease, particularly during in the early stages when
testing infrastructure, vaccines, and other medical interventions may
not be available at scale. Using a multiplex epidemic model that
consists of an information layer (modeling information exchange
between individuals) and a spatially embedded epidemic layer
(representing a human contact network), we study how random and
targeted disruptions in the information layer (\eg, errors and
intentional attacks on communication infrastructure) impact outbreak
dynamics. We calibrate our model to the early outbreak stages of the
SARS-CoV-2 pandemic in 2020. Mitigation campaign can still be
effective under random disruptions, such as failure of information
channels between a few individuals.  However, targeted disruptions or
sabotage of hub nodes that exchange information with a large number of
individuals can abruptly change outbreak characteristics such as the
time to reach the peak infection. Our results emphasize the importance
of using a robust communication infrastructure that can withstand both
random and targeted disruptions.
\end{abstract}

\maketitle
\begin{quotation}
Online communication platforms and exposure notification apps can help
slow down and contain the spread of an infectious
disease\cite{schneider2022epidemic}. Individuals who have been made
aware of an outbreak are likely to adapt their behavior to reduce
their risk of being infected. To study the interplay between
infectious disease outbreaks and corresponding changes in individual
contact behaviors, Granell \textit{et al.}\cite{granell2013dynamical}
introduced an epidemic model that accounts for the spread of awareness
through an information layer that is coupled to a human contact
network. Building upon their model of awareness diffusion, our work
studies the impact of random and targeted disruptions in the
information layer on the overall outbreak dynamics.
\end{quotation}
\section{Introduction}
\label{sec:intro}
The study of epidemic processes in networks has provided many insights
into the interplay between structure and
dynamics.\cite{gleeson2013binary,pastor2015epidemic} The aim of many
works in this area has been to analyze the impact of different
structural features such as clustering~\cite{newman2003properties},
community structure\cite{huang2007epidemic,tunc2014effects}, hub
nodes, and scale-free degree distributions~\cite{pastor2001epidemic}
on the evolution of susceptible-infected-susceptible (SIS) and
susceptible-infected-recovered (SIR) models and their
extensions.\cite{keeling2011modeling,bottcher2016connectivity,bottcher2018dynamical}
Connections between epidemic processes and percolation contributed to the development of analytical methods that are useful to analyze epidemic transitions and determine outbreak size.\cite{newman2002spread,d2019explosive,d2015anomalous,bottcher2020unifying,fan2020universal} Along with progress
in understanding epidemic processes in static single-layer networks,
developments in the study of temporal
networks~\cite{holme2012temporal}, multilayer
networks~\cite{de2013mathematical,kivela2014multilayer}, and other
structures describing higher-order
interactions~\cite{berge1973graphs,courtney2016generalized,
  majhi2022dynamics,anwar2022intralayer} have allowed for the
integration of time-varying and non-binary interactions.

Before research turned to epidemic models in multilayer networks,
interactions between disease and behavioral dynamics have been studied
mainly in single-layer networks\cite{wang2015coupled} and well-mixed
populations.\cite{fenichel2011adaptive,chen2012mathematical,
  reluga2013equilibria,bottcher2017critical} In an extension of the
classical SIS model, the so-called
susceptible-infected-alert-susceptible (SIAS) model, a new compartment
was used to study the effect of ``alert'' individuals that are
surrounded by a certain number of infecteds on disease
dynamics.\cite{DBLP:conf/cdc/SahnehS11,DBLP:conf/cdc/SahnehS12} The
SIAS model has been implemented using a two-layer
network~\cite{shakeri2015optimal} with a contact layer and an
information-dissemination layer to find optimal information
dissemination strategies that help contain an outbreak.

The interplay between behavioral effects and network dynamics has also
been analyzed in terms of a multiplex structure where information on
an outbreak diffuses in an information
layer.\cite{granell2013dynamical,granell2014competing} In a multiplex
network, all of the interlayer edges are edges between nodes and their
counterparts in other layers. As in the SIAS model, individuals in the
information layer can be either aware or unaware of a
disease. Awareness then translates into a reduced infection rate. The
original awareness model has been modified in various ways. One study
used a threshold model in the information layer and identified
awareness cascades.\cite{guo2015two} Other research investigated the
effects of dynamically varying transmission
rates~\cite{sagar2018effect}, coupled SIR and unaware-aware-unaware
(UAU) dynamics with and without
latency\cite{scata2020dynamical,wang2019impact}, SIS and UAU dynamics
that propagate at different speeds~\cite{velasquez2020disease}, and
higher-order interactions~\cite{fan2022epidemics}. For a detailed
overview of models of coevolving spreading processes in networks, we
refer the reader to Ref.~\citen{wang2019coevolution}.

In this work, we study coevolving
susceptible-exposed-infected-recovered-deceased (SEIRD) and UAU
dynamics on a multiplex network that consists of an epidemic layer and
an information layer. The exposed compartment in our model accounts
for latency (\ie, the time difference between infection and becoming
infectious). Different variants of SEIRD models have been used to
mechanistically describe the spread of an infectious disease for which
the latency period between time of infection to time of becoming
infectious cannot be neglected \cite{keeling2002understanding,
  keeling2011modeling,elderd2013population,bottcher2021decisive}.
Examples of such infectious diseases include measles, smallpox, and
SARS-CoV-2.

One of the main goals of this work is to provide insight into the
impact of disruptions in the information diffusion layer on the
overall outbreak dynamics. We therefore study different edge removal
protocols that describe random and targeted disruptions. In
Sec.~\ref{sec:methods}, we define the disease and awareness model,
develop a heterogeneous mean-field model, define random and targeted
edge removal protocols, and briefly describe the structure of the
considered networks. In Sec.~\ref{sec:results}, we first discuss a
baseline simulation that uses model parameters that are aligned with
empirical data on the outbreak of SARS-CoV-2 in early 2020. We then
use this baseline simulation as a reference to study the impact of
disruptions in the information diffusion layer on three disease
severity measures: (i) final outbreak size, (ii) maximum proportion of
infectious nodes on a given day (\ie, the height of the infection
peak), and (iii) the time until the infection peak is
reached.
%
\section{Methods}
\label{sec:methods}

\subsection{Epidemic model with information diffusion}
\begin{figure*}
	\centering
	\includegraphics[width=0.9\textwidth]{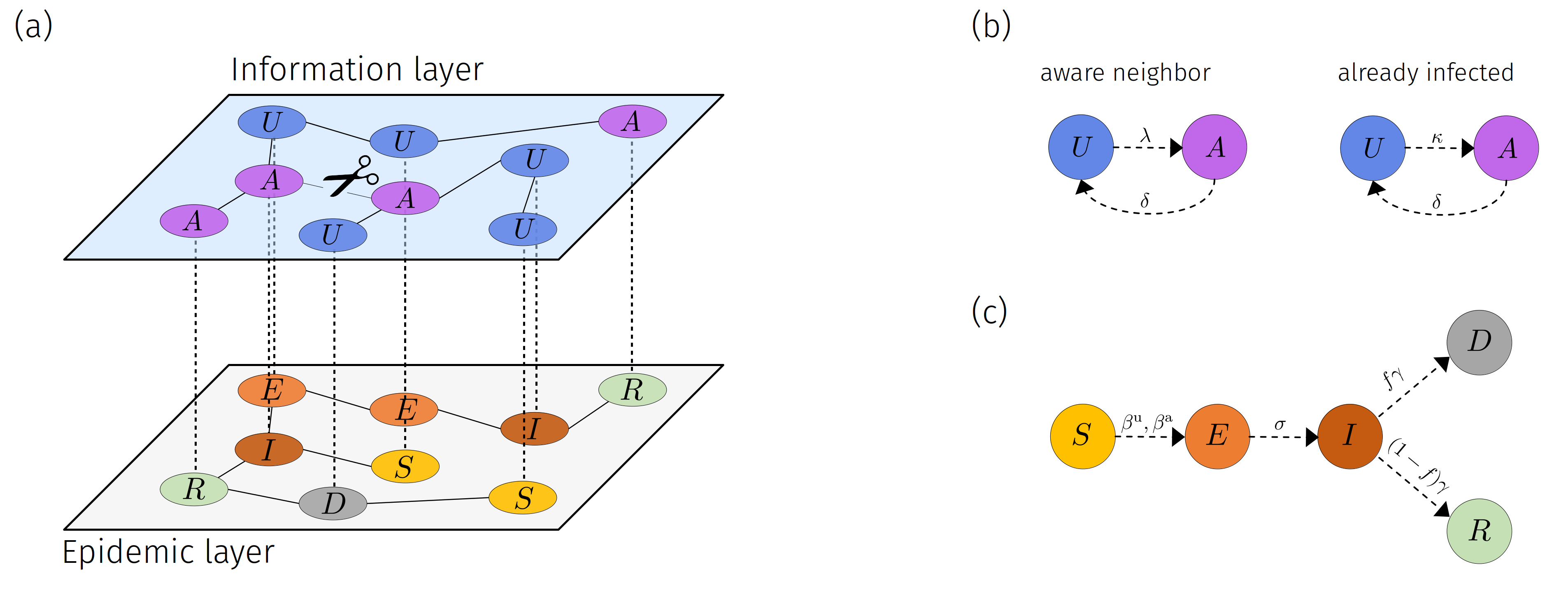}
	\caption{Model schematic. (a) Information layer and epidemic
          layer. Nodes in the information layer are either unaware
          ($U$) or aware ($A$) while nodes in the epidemic layer can
          be in one of five different states: susceptible ($S$),
          exposed ($E$), infected ($I$), recovered ($R$), and deceased
          ($D$). Edge removal that is caused by disruptions in the
          information layer is indicated by the scissor symbol. (b)
          Unaware nodes become aware at rate $\lambda$ if they are
          adjacent to an aware node. If unaware nodes are infected,
          they can also become aware at rate $\kappa$. Aware nodes
          transition back to an unaware state at rate $\delta$. (c)
          Infectious nodes transmit a disease to unaware and aware
          susceptible nodes at rates $\beta^{\rm u}$ and $\beta^{\rm
            a}$, respectively. To account for a reduction in
          infectiousness risk of aware nodes, we assume the value of
          the disease transmission rate $\beta^{\rm u}$ associated
          with unaware nodes is strictly larger than the value of the
          disease transmission rate $\beta^{\rm a}$ associated with
          aware nodes ($\beta^{\rm u}>\beta^{\rm a}$). Once
          susceptible nodes have been infected, they enter an exposed
          state and become infectious at rate $\sigma$. The
          characteristic time scale $\sigma^{-1}$ corresponds to the
          latency period of the disease. Infected nodes either die or
          recover at rates $f\gamma$ and $(1-f)\gamma$, respectively.}
	\label{fig:model_schematic}
\end{figure*}
We study the interplay between information diffusion and epidemic
dynamics in a multiplex network with two layers [see
  Fig.~\ref{fig:model_schematic}(a)].

In the first layer, individuals exchange information (\eg, through
online social media or messaging services) on the prevalence of a
certain disease in the overall population according to the
unaware-aware-unaware (UAU) model.\cite{granell2013dynamical}
Individuals in the ``information layer'' (IL) can be in two states. They are either unaware ($U$)
or aware ($A$) of the disease and do not necessarily have to be in
close proximity (in terms of connectivity) to exchange information. Unaware nodes can become aware in two
ways. First, if an unaware node is in contact with an aware node, it
becomes aware at rate $\lambda$. Second, nodes that have been infected
and experience symptoms become aware at rate $\kappa$. Given that
certain individuals forget or do not adhere to intervention measures
after a certain time, we also account for transitions from aware to
unaware at rate $\delta$. A schematic of UAU dynamics is shown in
Fig.~\ref{fig:model_schematic}(b).

In the second layer, we model an epidemic outbreak using the
susceptible-exposed-infected-recovered-deceased (SEIRD) model. In the
``epidemic layer'' (EL), nodes can be in states $S$ (susceptible), $E$
(exposed), $I$ (infected), $R$ (recovered), and $D$ (deceased). We
distinguish between two infection rates, $\beta^{\rm u}$ and
$\beta^{\rm a}$, that describe the rates at which susceptible nodes
become infected if they are unaware and aware, respectively. The
disease transmission rate associated with aware individuals is assumed
to be strictly lower than the disease transmission rate associated
with unaware individuals (\ie, $\beta^{\rm a}<\beta^{\rm u}$),
accounting for the decreased likelihood of an aware individual to
become infected. We assume a latent rate $\sigma$, resolution rate
$\gamma$, and infection fatality ratio $f$ that are independent of the awareness
status. This assumption is valid for infectious diseases for which no
medication is available that positively affects recovery, even if a
person is aware of an infection before developing symptoms. For
example, during the early outbreak stages of SARS-CoV-2, there was
very little information available on how to medically support patients
that were aware of their infection, but did not show symptoms
yet. Non-pharmaceutical interventions such as contact restrictions,
mask mandates, and quarantine are often the only possibility to combat
novel pathogens.\cite{schneider2022epidemic}

According to the described UAU and SEIRD dynamics, nodes can be in the
following states: $(U,S)$, $(A,S)$, $(U,E)$, $(A,E)$, $(U,I)$,
$(A,I)$, $(U,R)$, $(A,R)$, and $(U,D)$. The first entry in each tuple
describes the awareness state (either $U$ or $A$) while the second
entry describes vital and disease states ($S$, $E$, $I$, $R$, and
$D$). Deceased nodes are not aware.
\subsection{Heterogeneous mean-field theory}
In accordance with Ref.~\citen{zhou2019effective}, we formulate a
heterogeneous mean-field theory of SEIRD-UAU dynamics. We use $x_j
y_k\equiv x_j y_k (t)$ ($x\in\{u,a\},y\in\{s,e,i,r,d\}$) to denote the
proportion of nodes in state $X_j Y_k$
($X\in\{U,A\},Y\in\{S,E,I,R,D\}$) with degrees $j$ and $k$ in the IL
and EL at time $t$, respectively. For example, $u_j s_k\equiv u_j
s_k(t)$ denotes the proportion of unaware and susceptible nodes with
degrees $j$ and $k$ in the IL and EL at time $t$, respectively.
Henceforth, we will not explicitly include the time dependence in the
notation $x_j y_k$ for the sake of notational brevity.

The proportions of susceptible, exposed, infected, recovered, and
deceased nodes are
\begin{align}
s_{k}&=\sum_{j=1}^{J}(u_j s_{k}+a_j s_{k})\,,\\
e_{k}&=\sum_{j=1}^{J}(u_j e_{k}+a_j e_{k})\,,\\
i_{k}&=\sum_{j=1}^{J}(u_j i_{k}+a_j i_{k})\,,\\
r_{k}&=\sum_{j=1}^{J}(u_j r_{k}+a_j r_{k})\,,\\
d_{k}&=\sum_{j=1}^{J}u_j d_k\,,
\end{align}
where $J$ is the maximum (or cut-off) degree in the IL. Similarly, we
find that the proportions of unaware and aware nodes are
\begin{align}
u_j &= \sum_{k=1}^{K} (u_j s_{k}+u_j e_{k}+u_j i_{k}+u_j r_{k}+d_k)\,,\\
a_j &= \sum_{k=1}^{K} (a_j s_{k}+a_j e_{k}+a_j i_{k}+a_j r_{k})\,,
\end{align}
where $K$ is the maximum (or cut-off) degree in the EL. These
quantities satisfy the normalization conditions
\begin{align}
&\sum_{k=1}^{K} (s_k+e_k+i_k+r_k+d_k)=1\,,\\ &\sum_{j=1}^{J} (u_j +
  a_j) = 1\,.
\end{align}

Assuming an uncorrelated network~\cite{xia2022controlling}, the rate
equations of the heterogeneous mean-field model are
\begin{align}
  \frac{\mathrm{d}u_js_k}{\mathrm{d}t} = & -\lambda \frac{j
    u_js_k}{\langle \tilde{k}\rangle} \sum_{j'} j' a_{j'} -\beta^{\rm
    u} \frac{k u_js_k}{\langle k\rangle} \sum_{k'} k' i_{k'} +\delta
  a_js_k\,,\label{eq:dusdt}\\
  \frac{\mathrm{d}a_js_k}{\mathrm{d}t} = &  \lambda \frac{j
    u_js_k}{\langle \tilde{k}\rangle} \sum_{j'} j' a_{j'}-\beta^{\rm
    a} \frac{k a_js_k}{\langle k\rangle} \sum_{k'} k' i_{k'}-\delta
  a_js_k \,,\\
  \frac{\mathrm{d}u_je_k}{\mathrm{d}t} = & -\lambda \frac{j
    u_je_k}{\langle \tilde{k}\rangle} \sum_{j'} j' a_{j'}+\beta^{\rm
    u} \frac{k u_js_k}{\langle k\rangle} \sum_{k'} k' i_{k'}\\
\: & -\sigma u_je_k+\delta a_je_k\nonumber\,
\end{align}
and
\begin{align}
\frac{\mathrm{d}a_je_k}{\mathrm{d}t} = & \lambda \frac{j
  u_je_k}{\langle \tilde{k}\rangle} \sum_{j'} j' a_{j'}+\beta^{\rm a}
\frac{k a_js_k}{\langle k\rangle} \sum_{k'} k' i_{k'}\\ 
\: & -\sigma a_j e_k-\delta a_j e_k\,,\nonumber\\
\frac{\mathrm{d}u_ji_k}{\mathrm{d}t} = & -\lambda \frac{j
  u_ji_k}{\langle \tilde{k}\rangle} \sum_{j'} j' a_{j'}+\sigma
u_je_k-\gamma u_ji_k\\
\: & -\kappa u_j i_k +\delta a_ji_k\,,\nonumber\\
\frac{\mathrm{d}a_ji_k}{\mathrm{d}t} = & \lambda \frac{j u_ji_k}{\langle
  \tilde{k}\rangle} \sum_{j'} j' a_{j'}+\sigma a_je_k-\gamma a_ji_k\\
\: & +\kappa u_j i_k-\delta a_ji_k\,,\nonumber\\
\frac{\mathrm{d}u_jr_k}{\mathrm{d}t} = & -\lambda \frac{j u_j
  r_k}{\langle \tilde{k}\rangle} \sum_{j'} j' a_{j'}+(1-f)\gamma
u_j i_k +\delta a_j r_k\,,\\
\frac{\mathrm{d}a_jr_k}{\mathrm{d}t} = & \lambda \frac{j u_j
  r_k}{\langle \tilde{k}\rangle} \sum_{j'} j' a_{j'}+(1-f)\gamma
a_ji_k-\delta a_jr_k\,,\\
\frac{\mathrm{d}u_j d_k}{\mathrm{d}t} = & f\gamma (u_j+a_j)i_k\label{eq:duddt}\,,
\end{align}
where $\langle k\rangle$ and $\langle \tilde{k} \rangle$ denote the
mean degrees of the EL and IL, respectively.
\subsection{Networks}
\begin{figure*}
	\centering
	\includegraphics[width=0.45\textwidth]{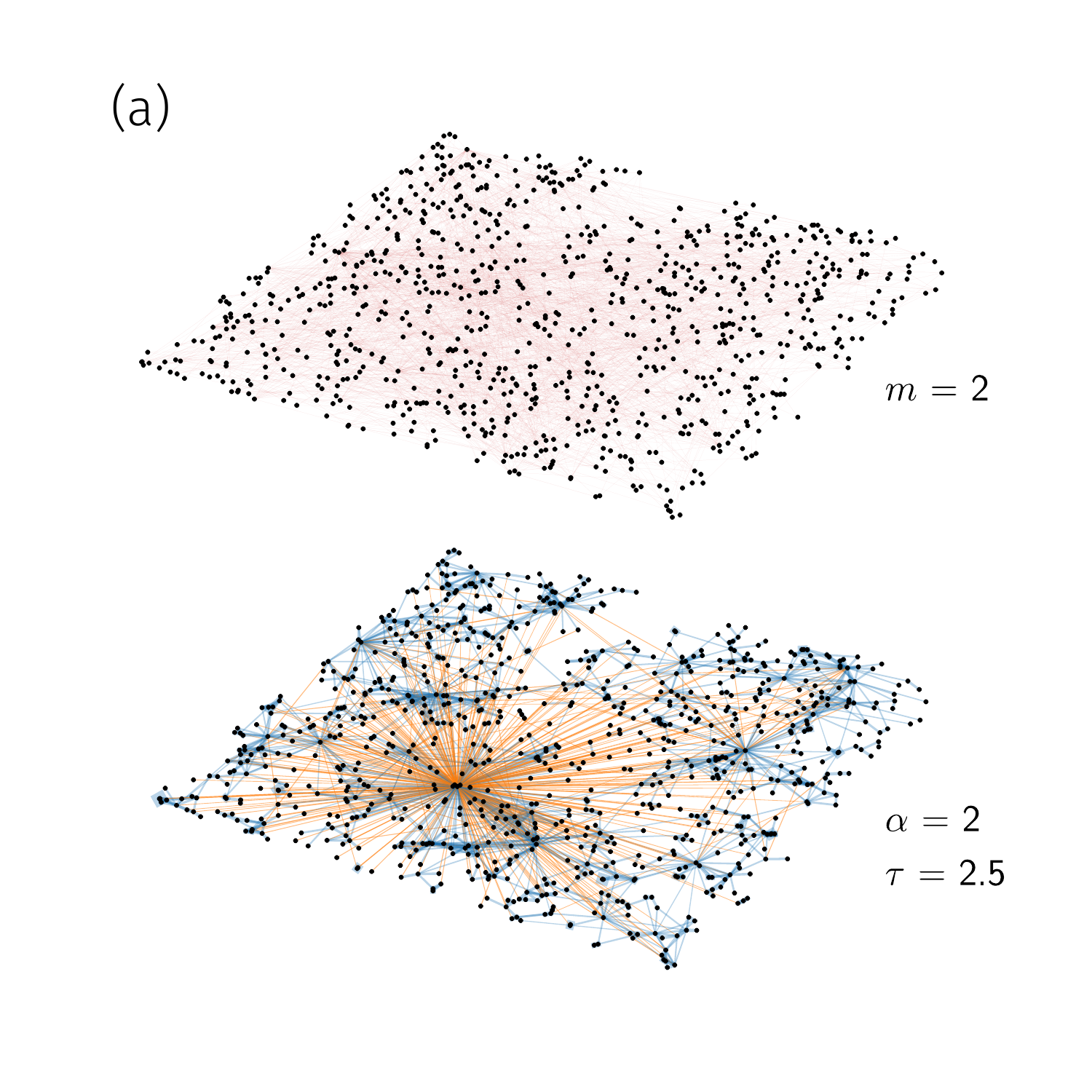}
	\includegraphics[width=0.45\textwidth]{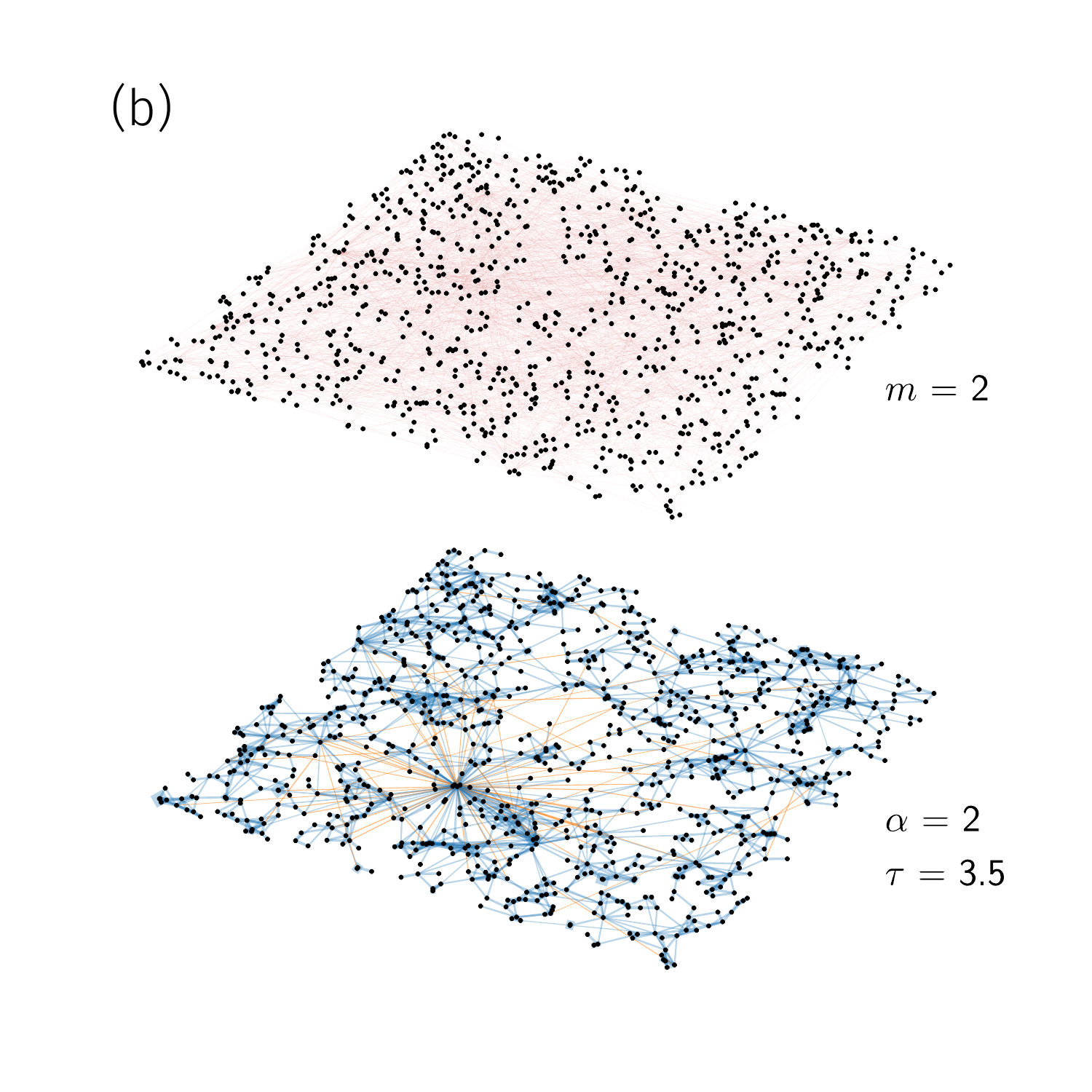}
	\caption{Multiplex networks. Information layer (top layer)
          with BA structure and and epidemic layer (bottom layer) with
          GIRG structure determined by exponents $\alpha=2$,
          $\tau=2.5$ (a) and $\alpha=2$, $\tau=3.5$ (b). In the BA
          network, each new node has $m=2$ edges that connect it to
          existing nodes using linear preferential attachment. We use
          blue and orange edges in the epidemic layer to indicate
          short-range and long-range connections, respectively. An
          edge connecting two nodes $i,j$ is considered a short-range
          connection if the corresponding positions
          $\mathbf{x}_i,\mathbf{x}_j$ satisfy $\|\mathbf{x}_{i} -
          \mathbf{x}_{j}\| < 7$. Otherwise, it is considered a
          long-range connection. The numbers of nodes in panels (a)
          and (b) are $N=921$ and $N=973$, respectively.}
	\label{fig:BA-GIRG}
\end{figure*}
In our numerical experiments, we use a Barabási--Albert (BA)
network~\cite{barabasi1999emergence} to model the information layer of
the two-layer structure underlying SEIRD-UAU dynamics. Such networks
exhibit scale-free degree distributions $p(k)\propto k^{-\gamma}$
($\gamma>0$), and are often found in social and technological
systems.\cite{ebel2002scale, barabasi2002evolution,
  albert1999diameter,danon2012social,brown2013place} Note that other
distributions such as log-normal distributions may also provide good
descriptions of empirical degree distributions in seemingly scale-free
networks.\cite{broido2019scale} In the epidemic layer, we use a
geometric inhomogeneous random graph
(GIRG)~\cite{bringmann2019geometric}, a spatial network that has found
applications in representing spatially embedded metapopulation
structures in COVID-19 models.\cite{odor2021switchover}
\subsubsection{Barabási--Albert network}
\label{sec:BA}
Barabási--Albert networks~\cite{barabasi1999emergence} are constructed
using a preferential attachment procedure in which new nodes that are
iteratively added to an existing network have a higher likelihood of
being attached to nodes that have higher numbers of connections. A
mean field analysis of the BA model and corresponding numerical
results show that the exponent of the power-law degree distribution is
$\gamma \approx 3$.\cite{barabasi1999mean}

To construct the BA network that we will use in our simulations, we start
with a star graph with one root node and two leave nodes and iteratively add new nodes until we reach $N$
nodes. Each new node has $m=2$ edges that connect it to existing nodes
using linear preferential attachment. A visualization of such a BA
information layer network with $N\approx 10^3$ is given in the top row of
Fig.~\ref{fig:BA-GIRG}. In our simulations, we use a BA network with a larger node number of $N\approx 10^4$ that is constructed in the same way as the ILs in Fig.~\ref{fig:BA-GIRG}.
\subsubsection{Geometric inhomogeneous random graph}
\label{sec:GIRG}
The GIRG model\cite{bringmann2019geometric,joostjor_2020} produces a spatially
embedded scale-free random network. In this model, $N$ points are
first selected uniformly at random in the $n$-dimensional hypercube
$K^{n}=[0,1]^{n}$. We denote the randomly selected point positions by
$\mathbf{x}_{i} \in K^{n}$ ($1\leq i\leq N$) and assign each of them a
weight $w_{i}$ whose value is drawn from a power-law distribution $\tilde{p}(w) =(\tau-2) w^{-\tau}$ ($w\geq 1,\tau \geq 2$).\cite{bringmann2019geometric,joostjor_2020} Note that the distribution $\tilde{p}(w)$ is normalized such that its mean value is equal to 1. Pairs of nodes $i,j$ with positions $\mathbf{x}_{i}, \mathbf{x}_{j}$ are adjacent with probability
\begin{equation}
	\Pi_{ij} = 1 - \exp \left[-\left( \frac{ w_{i} ~ w_{j}} 
{\|\mathbf{x}_{i} - \mathbf{x}_{j}\|^{n} } \right)^{\alpha} \right]\,,
\label{eqn:GIRG}
\end{equation}
where $\|\mathbf{x}_{i} - \mathbf{x}_{j}\|$ denotes the Euclidean
distance between points $\mathbf{x}_{i}$ and $\mathbf{x}_{j}$. The resulting degrees $k_i$ ($1\leq i\leq N$) are also distributed according to a power law with exponent $\tau$.

According to Eq.~\ref{eqn:GIRG}, the exponent $\alpha$ tunes the
distance and weight dependence of $\Pi_{ij}$. For $\alpha=0$, the
probability that two nodes $i,j$ are adjacent is independent of their
distance $\vert \mathbf{x}_i - \mathbf{x}_j\vert$. That is,
$\Pi_{ij}=1-{\rm e}^{-1}$ for all $i,j$. By increasing $\alpha$, the
distance-dependence of $\Pi_{ij}$ strongly influences the structure of
the network so that only nearby nodes are likely to be adjacent. The
bottom row of Fig.~\ref{fig:BA-GIRG} shows GIRGs for various
parameters.

For small exponents $\tau\geq 2$, the number of nodes with large weight
values increases. According to Eq.~\ref{eqn:GIRG}, nodes with large
weights are more likely to be connected than nodes with small
weights. The abundance of these large-weight nodes, which are the hubs
of the underlying scale-free network, impacts the global structure of
GIRG. By decreasing $\tau$, many long-range connections are added to
the GIRG. In the bottom row of Fig.~\ref{fig:BA-GIRG}, we observe that
smaller values of $\tau$ are associated with a larger proportion of
long-range connections.
\subsection{Edge removal}
\label{sec:edge_removal}
\begin{table*}
	\centering
	\renewcommand*{\arraystretch}{1.5}
	\begin{tabular}{lcccc}\toprule
		\multicolumn{1}{c}{Parameter} & \multicolumn{1}{c}{Symbol} & \makebox[6em]{Value} & \multicolumn{1}{c}{Units} & \multicolumn{1}{c}{Comments/references} \\\hline
		\,\,\,Infection rate (unaware) \,\,\, & $\beta^{\rm u}$ & $0.17, 0.6$ & $\mathrm{day}^{-1}$ & \,\,\, inferred from $R_0\approx2-4$ for a given $\gamma$~\cite{lai2020severe,park2020reconciling}\,\,\, \\
		\,\,\,Infection rate (aware) \,\,\, & $\beta^{\rm a}$ & $0.2\beta^{\rm u}$ & $\mathrm{day}^{-1}$ & \,\,\, \cite{liu2021rapid}\,\,\, \\
		\,\,\,Latent rate \,\,\, & $\sigma$ & $1/5$ & $\mathrm{day}^{-1}$ & \,\,\, \cite{xin2022estimating}\,\,\, \\
		\,\,\,Resolution rate \,\,\, & $\gamma$ & $1/14$ & $\mathrm{day}^{-1}$ & \,\,\, \cite{bottcher2020case,CDCrecovery}\,\,\, \\
		\,\,\,Infection fatality ratio \,\,\, & $f$ & $1\%$ & $\dots$ & \,\,\, \cite{salje2020estimating,bottcher2021using} \,\,\, \\
		\,\,\,Awareness rate (infected)\,\,\, & $\kappa$ & $1$ & $\mathrm{day}^{-1}$ & \,\,\, \cite{teslya2020impact} \,\,\, \\
		\,\,\,Base awareness rate \,\,\, & $\lambda$ & $0.5\kappa$ & $\mathrm{day}^{-1}$ & \,\,\, \cite{teslya2020impact} \,\,\, \\
		\,\,\,Unawareness rate \,\,\, & $\delta$ & $1/30$ & $\mathrm{day}^{-1}$ & \,\,\, \cite{teslya2020impact} \,\,\, \\\bottomrule
	\end{tabular}
	\vspace{1mm}
	\caption{Overview of model parameters. We use infection rates
          $\beta^{\rm u}=0.17~\mathrm{day}^{-1}$ and $\beta^{\rm
            u}=0.6~\mathrm{day}^{-1}$ for GIRG networks with
          $\tau=2.5$ (long range) and $\tau=3.5$ (short range),
          respectively.}
	\label{tab:parameters}
\end{table*}
To model disruptions in the IL, we consider two different edge-removal
protocols: (i) random edge removal and (ii) targeted edge removal. In
both protocols, we select $\tilde{N}\leq N$ nodes and denote the
proportion of selected nodes by $q=\tilde{N}/N$. For each selected
node, we remove each of its edges with probability $p$.  Values of
$p,q>0$ correspond to disruptions in the IL that slow down the
information spread. For $p=q=1$, there are no awareness dynamics and
the epidemic progresses without interference from the information
layer.

In random edge removal, $\tilde{N}$ nodes are selected uniformly at
random while we select $\tilde{N}$ hub nodes (\ie, nodes with the
largest degrees) in targeted edge removal.  Such random and targeted
disruptions have been studied to provide insight into the ability of
different types of networks to withstand errors and intentional
attacks.\cite{albert2000error} It has been shown that structural
features of scale-free networks such as the size of the largest
connected component are very sensitive to intentional attacks (or
sabotage).\cite{cohen2000resilience,cohen2001breakdown}

We next explore how variations in $p,q \in [0,1]$ impact the total
proportion of infections $i^*=1-s^*$, peak infection (\ie, the maximum
proportion of the population that was infected on any day), and the
time between the beginning of the outbreak until peak infection is
reached.
\section{Results}
\label{sec:results}
First consider a baseline case of SEIRD-UAU dynamics without edge
removal (\ie, $p q=0$) in two different multiplex networks. Both
multiplex networks are connected and have the same BA information
layer (see Sec.~\ref{sec:BA}). In the epidemic layer, we set
$\tau=3.5$ and $\tau=2.5$ to model contact networks with different
proportions of long-range connections (see Fig.~\ref{fig:BA-GIRG}). In
the remainder of this work, we will refer to the networks with
$\tau=2.5$ and $\tau=3.5$ as long-range and short-range networks,
respectively. In both networks, we set $\alpha=2$ [see
  Eq.~\eqref{eqn:GIRG}]. All stochastic simulations are implemented
using Gillespie's
algorithm.\cite{gillespie1976general,gillespie1977exact,bottcher2021computational}
\subsection{Baseline}
\begin{figure}
    \centering
    \includegraphics{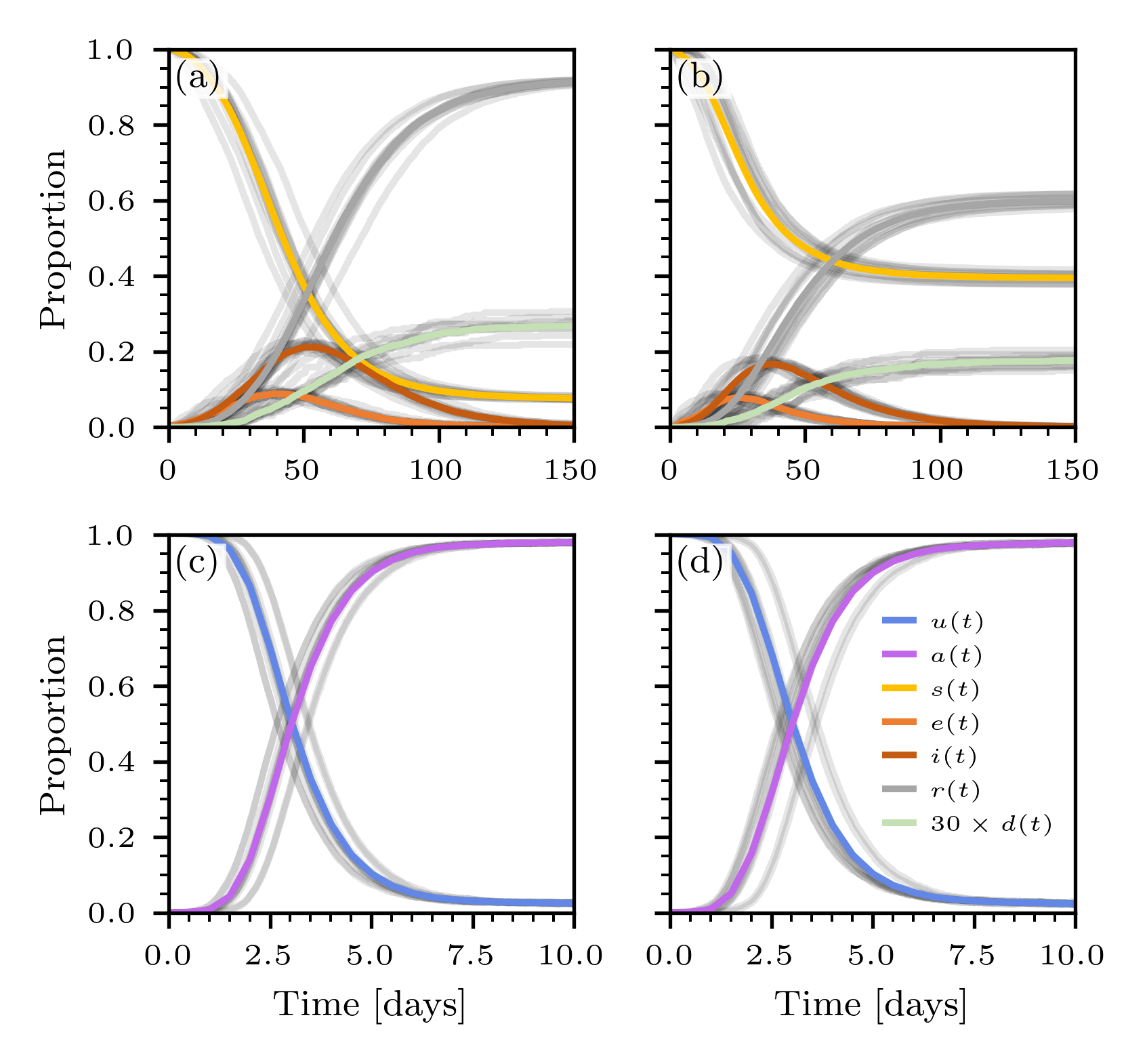}
    \caption{Stochastic simulation of baseline scenario without
      information-layer disruption (\ie, $pq=0$). (a,b) Proportions of
      susceptible ($s(t)$), exposed ($e(t)$), infected ($i(t)$),
      recovered ($r(t)$), and deceased ($d(t)$) nodes at time $t$. The
      exponent $\tau$ in the epidemic layer in panels (a,c) and (b,d)
      is set to $3.5$ (short range) and $2.5$ (long range),
      respectively. The corresponding numbers of nodes are $N=10049$
      and $N=10025$. Solid colored lines represent mean values that
      are based on 10 i.i.d.\ realizations (thin grey lines).}
    \label{fig:baseline}
\end{figure}
\begin{figure}
    \centering
    \includegraphics{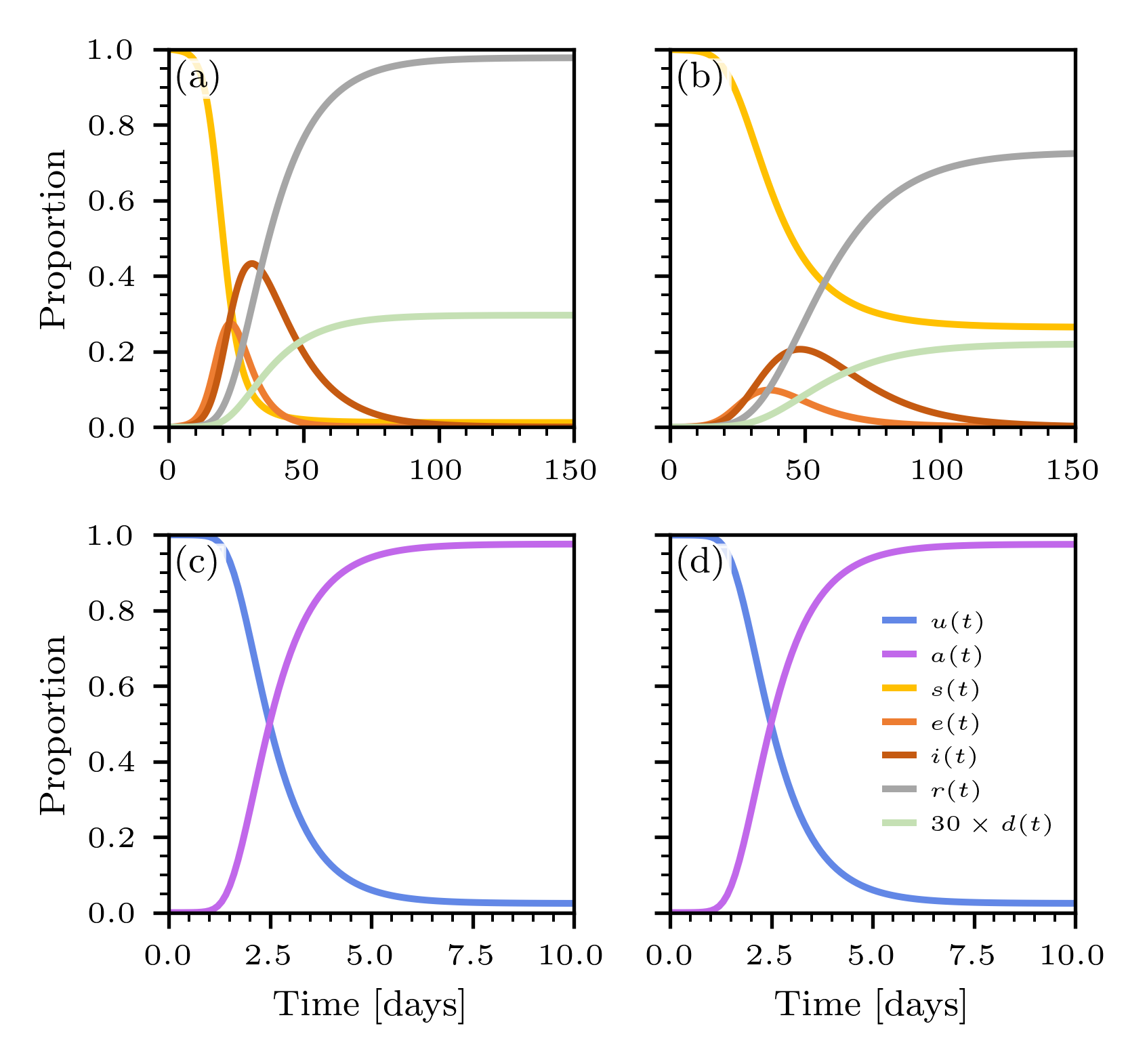}
    \caption{Heterogeneous mean-field solution of baseline scenario
      without information-layer disruption (\ie, $pq=0$). (a,b)
      Proportions of susceptible ($s(t)$), exposed ($e(t)$), infected
      ($i(t)$), recovered ($r(t)$), and deceased ($d(t)$) nodes at
      time $t$. The exponent $\tau$ in the epidemic layer in panels
      (a,c) and (b,d) is set to $3.5$ (short range) and $2.5$ (long
      range), respectively. The corresponding numbers of nodes are
      $N=10049$ and $N=10025$.}
    \label{fig:baseline_mean_field}
\end{figure}
We have chosen the model parameters that we use in the baseline
simulation in accordance with empirical data on the outbreak of
SARS-CoV-2 in the beginning of 2020. For example, for the two
multiplex networks that we use in our simulations, we have set the
infection rate of unaware nodes to $\beta^{\rm
  u}=0.17,0.6~\mathrm{day}^{-1}$ to obtain a basic reproduction number
$R_0$ of about $2-4$.\cite{lai2020severe,park2020reconciling} Given a
latency period of about 5 days\cite{xin2022estimating}, we set the
latent rate to $\sigma=1/5~\mathrm{day}^{-1}$. The resolution rate is
set to $\gamma=1/14~\mathrm{day}^{-1}$, and we use an infection fatality ratio
$f$ of
1\%.\cite{bottcher2020case,CDCrecovery,salje2020estimating,bottcher2021using}
Other model parameters that are associated with UAU dynamics are as in
Ref.~\citen{teslya2020impact}. We provide an overview of all
parameters and corresponding references in Tab.~\ref{tab:parameters}.

Figure~\ref{fig:baseline} shows the stochastic evolution of the proportions of
susceptible $s(t)$, exposed $e(t)$, infected $i(t)$, recovered $r(t)$,
and deceased $d(t)$ nodes in the EL and of unaware $u(t)$ and aware
$a(t)$ nodes in the IL. Initially, 10 nodes are infectious and 1 node
is aware. For networks of about $N=10000$ nodes that are used in our
stochastic simulations, these initial conditions correspond to $i(0)\approx 10^{-3}$ and $a(0)\approx
10^{-4}$. The simulation results shown in
Figs.~\ref{fig:baseline}(a,c) and Figs.~\ref{fig:baseline}(b,d) are
based on short-range ($\tau=3.5$) and long-range ($\tau=2.5$) GIRGs, respectively. The evolution of the UAU dynamics in the IL is
very similar for both GIRGs. However, structural differences between
the ELs directly impact the evolution of SEIRD dynamics. The infected
fraction peaks at $\sim 0.17$ after about 38 days in the long-range EL
but peaks at $\sim 0.21$ at about 51 days in the short-range EL. Figure~\ref{fig:baseline}
also shows that the final epidemic size $1-s(t\rightarrow\infty)$ in
both networks differs significantly. To understand what causes the
different outbreak characteristics in both networks, we examined the
degree distribution of susceptible nodes at $T=150$: there are
substantially more susceptible low-degree nodes in the long-range GIRG
where $\tau=2.5$ compared to the short-range GIRG with
$\tau=3.5$. Although, there are more hub nodes with large degree in
the long-range GIRG, the proportion of low-degree nodes is also
larger. Hence, there are more low-degree nodes in the long-range GIRG
that are less exposed to the outbreak dynamics.

To complement the stochastic simulation results, we numerically solve the
heterogeneous mean-field model \eqref{eq:dusdt}-\eqref{eq:duddt} for
the same networks and model parameters (see
Tab.~\ref{tab:parameters}). We set the degree cut-offs to $J=210$,
$K=400$ ($\tau=2.5$) and $J=210$, $K=164$ ($\tau=3.5$). In the
multiplex network with short-range IL with $\tau=3.5$, the degree
cut-offs correspond to the maximum degrees. In the long-range EL where
$\tau=2.5$, the maximum degree is 856, and to keep the solution of the
mean-field model computationally feasible we set the cut-off
$K=400$. Initially, we set $a_ji_k(0)=p_j\tilde{p}_k a(0)/2$,
$a_js_k(0)=p_j\tilde{p}_k a(0)/2$, $u_js_k(0)=p_j\tilde{p}_k
(1-i(0)-a(0)/2)$, $u_ji_k(0)=p_j\tilde{p}_k (i(0)-a(0)/2)$, where
$p_j$ and $\tilde{p}_k$ denote the degree distributions in the IL and
EL, respectively. Both degree distributions are normalized according
to $\sum_{j=1}^Jp_j=1$ and $\sum_{k=1}^K\tilde{p}_k=1$.
\begin{figure*}
    \centering
    \includegraphics[width=0.8\textwidth]{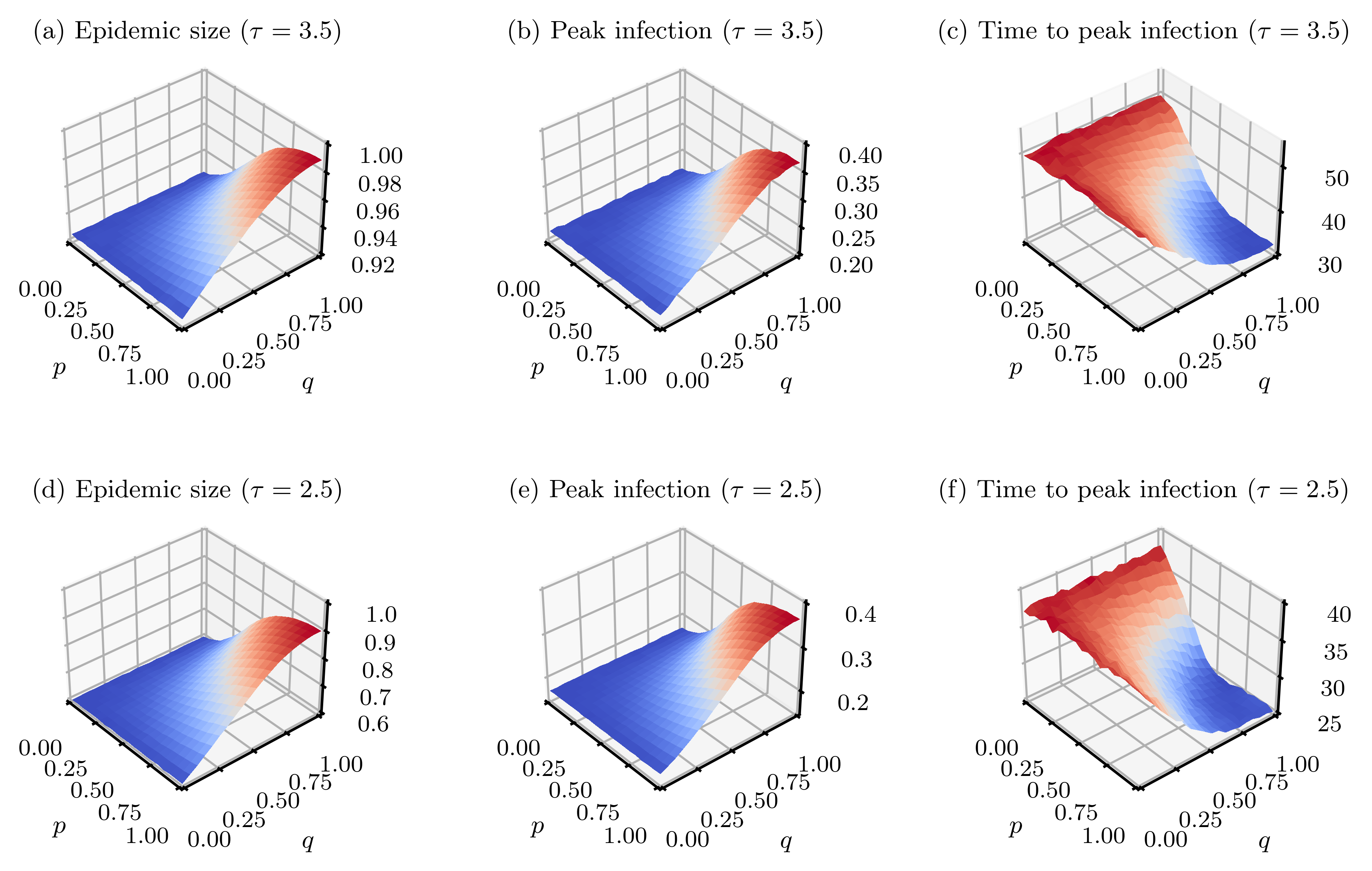}
    \caption{Random edge removal. The impact of random edge removal in
      the IL on disease dynamics in the EL. Epidemic size
      $1-s(t\rightarrow\infty)$ (left column), peak infection (middle
      column), and time to peak infection (right column) as a function
      of the proportion of selected nodes $q$ and the corresponding
      edge removal probability $p$. The exponent $\tau$ in the ELs in
      top row and bottom row is set to $3.5$ (short range) and $2.5$
      (long range), respectively. The corresponding numbers of nodes
      are $N=10049$ and $N=10025$. Simulation results are based on 230
      i.i.d.\ realizations.}
    \label{fig:edge_removal_random}
\end{figure*}
\begin{figure*}
    \centering
    \includegraphics[width=0.8\textwidth]{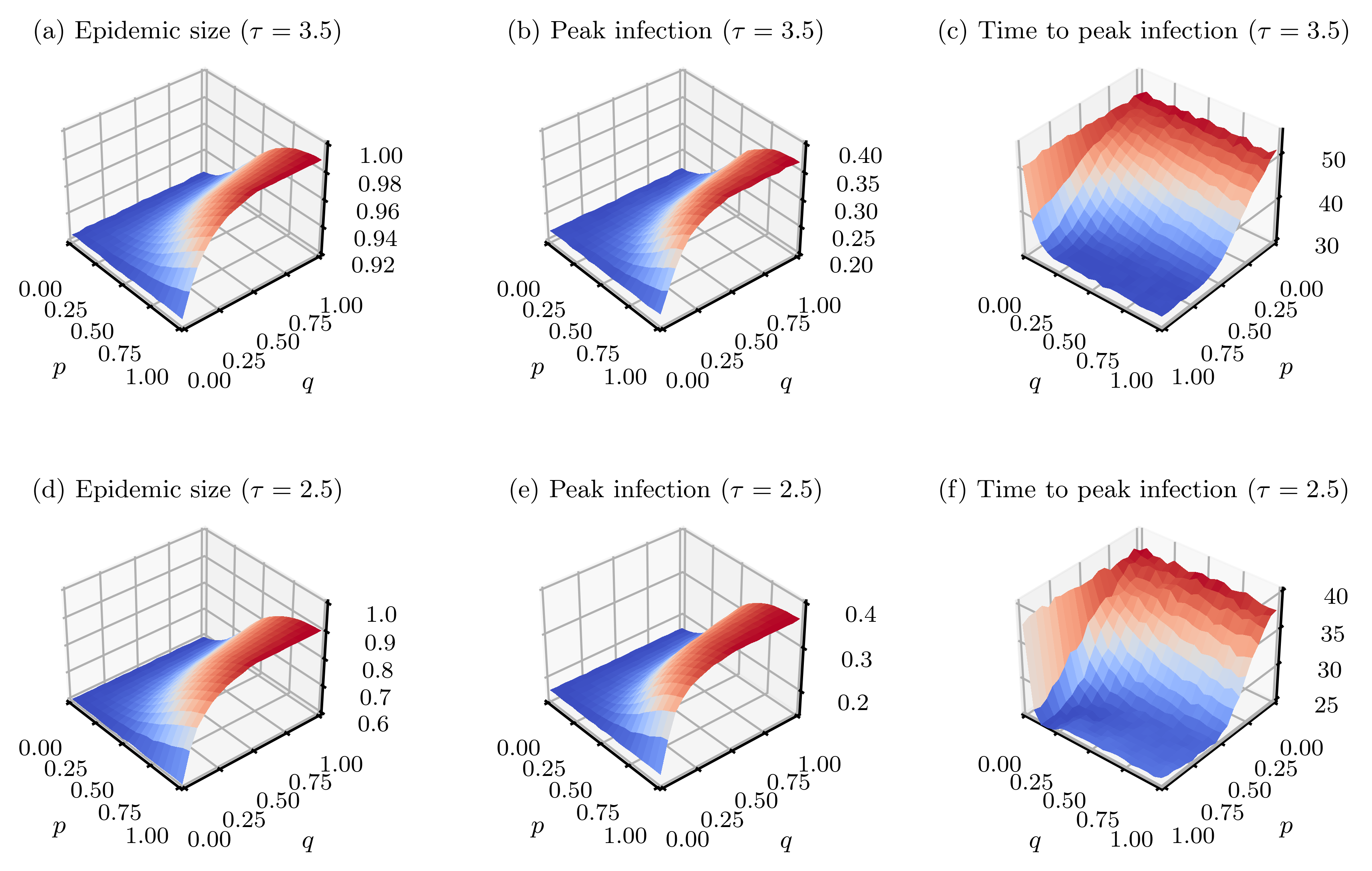}
    \caption{Targeted edge removal. The impact of random edge removal
      in the IL on disease dynamics in the EL. Epidemic size
      $1-s(t\rightarrow\infty)$ (left panel), peak infection (middle
      panel), and time to peak infection (right panel) as a function
      of the proportion of selected nodes $q$ and the corresponding
      edge removal probability $p$. The exponent $\tau$ in the ELs in
      top row and bottom row is set to $3.5$ (short range) and $2.5$
      (long range), respectively. The corresponding numbers of nodes
      are $N=10049$ and $N=10025$. Simulation results are based on 230
      i.i.d.\ realizations.}
    \label{fig:edge_removal_targeted}
\end{figure*}

Note that these initial conditions satisfy
\begin{align}
s(0) & =\sum_{j,k}\big(u_js_k(0)+a_js_k(0)\big) \\
\: &  =\sum_{j,k} p_j\tilde{p}_k\big[1-i(0)\big]=1-i(0)\,,\\
i(0) &=\sum_{j,k} \big(u_ji_k(0)+a_ji_k(0)\big) =\sum_{j,k} p_j\tilde{p}_k  i(0)\,,\\
u(0) &=\sum_{j,k} \big(u_js_k(0) + u_ji_k(0)\big)\\
\: &  =\sum_{j,k} p_j\tilde{p}_k\big[ 1-a(0)\big]=1-a(0)\,,\\
a(0) &=\sum_{j,k} \big(a_js_k(0) + a_ji_k(0)\big) =\sum_{j,k} p_j\tilde{p}_k  a(0)\,.
\end{align}
In accordance with the initial conditions that we used in the
stochastic simulations, we set $i(0)=10^{-3}$ and
$a(0)=10^{-4}$. Figure~\ref{fig:baseline_mean_field} shows the
corresponding numerical results. Comparing Figs.~\ref{fig:baseline} to
\ref{fig:baseline_mean_field}, we observe that the heterogeneous
mean-field model captures characteristic features that arise in the
evolution of stochastic SEIRD-UAU dynamics. Examples of such features
include (i) the rapid spread of awareness in the IL and (ii)
differences between both ELs in the final epidemic size $1-s(t\rightarrow\infty)$. In the heterogeneous mean-field model \eqref{eq:dusdt}-\eqref{eq:duddt}, we account only for differences in node degree and neglect other structural features of the considered multiplex networks. Subpopulations interact in a well-mixed manner and susceptible nodes of the same degree have the same risk of being infected at any given time. As a consequence of these approximations, the mean-field model overestimates both the number of new infections and final outbreak size compared to the stochastic simulation results in Fig.~\ref{fig:baseline}.
\subsection{Impact of edge removal}
We now study the impact of random and targeted edge removal in the IL
(see Sec.~\ref{sec:edge_removal}) on SEIRD dynamics in terms of three
disease severity measures: (i) final epidemic size, (ii) peak
infection, and (iii) time to peak infection.
\subsubsection{Random edge removal}
In random edge removal, we first select a proportion of
$q=\tilde{N}/N$ nodes in the IL uniformly at random. For each of the
selected nodes, each of its edges are removed with probability $p$.

Figure~\ref{fig:edge_removal_random}(a,d) shows the epidemic size as a
function of $p,q$ for both short-range and long-range GIRGs.  The
epidemic size increases with $p$ and $q$ because larger values of
$p,q$ are associated with fewer edges in the IL, leading to a smaller
proportion of aware nodes. Hence, the proportion of nodes with a
reduced infection rate $\beta^{\rm u}$ also decreases. For the
long-range GIRG ($\tau=2.5$), the final epidemic size undergoes a
transition from about 0.6 for $p,q \approx 0$ to about 0.9 for $p,q
\approx 1$. Because the final epidemic size in the short-range GIRG
(${\tau=3.5}$) is already about 0.9, random edge removal has relatively
little impact on this quantity.

As with the impact on final epidemic size $1-s(t\rightarrow\infty)$,
random edge removal generates a similar-looking $p,q$-dependent
infection peak, as shown in
Fig.~\ref{fig:edge_removal_random}(b,e). The time to reach peak
infection decreases with $p,q$ since higher $p,q$ are associated with
smaller proportions of aware nodes. Thus, the proportion of nodes with
a reduced infection rate $\beta^{\rm u}$ also decreases, and the
epidemic spreads faster through the network.
\subsubsection{Targeted edge removal}
For targeted edge removal where the $\tilde{N}$ selected nodes
correspond to the hubs (\ie, largest-degree nodes) of the IL, we find
that the overall dependence of epidemic size, peak infection, and time
to peak infection on $p,q$ is qualitatively similar to random edge
removal (see Fig.~\ref{fig:edge_removal_targeted}). As in random edge
removal, the impact of targeted edge removal on the final epidemic
size is smaller for the short-range GIRG compared to the long-range
one. A key difference in targeted edge removal is that all studied
quantities are more sensitive to variations in $q$, the proportion of
selected hub nodes. For example, the transition of the epidemic size
for $p=1$ as a function of $q$ in targeted edge removal [see
  Fig.~\ref{fig:edge_removal_targeted}(a,d)] is steeper than the
corresponding transition in random edge removal [see
  Fig.~\ref{fig:edge_removal_random}(a,d)].

Targeted edge removal selects nodes based on their degree rather than
uniformly, and leads to more significant changes in epidemic size,
peak infection, and time to peak infection as $p \geq 0.5$.  These
findings are in accordance with previous work that showed that
scale-free networks break down more easily under intentional attacks
than under uniform random failure.\cite{cohen2001breakdown} Our work
provide insights into how such disruptions in information diffusion
translate into differences in disease severity measures.
\section{Discussion}
\label{sec:discussion}
In this work, we studied the impact of disruptions in communication
networks on information diffusion and subsequently disease outcome
during an outbreak. To do so, we constructed a multiplex network that
consists of two layers. The first layer, called information layer
(IL), is used to model communication between individuals (\eg, online
information exchange via a social media platform). The second layer,
called epidemic layer (EL), is used to represent a spatially embedded
human contact network in which infectious individuals can transmit a
disease to susceptible individuals. We use this multiplex network to
simulate coevolving unaware-aware-unaware (UAU) and
susceptible-exposed-infected-recovered-deceased (SEIRD) dynamics. The
model parameters that we use in our simulations have been selected in
accordance with empirical data on the early outbreak stages of
SARS-CoV-2 in the beginning of 2020.

We studied two different epidemic layers with different proportions of
long-range connections, representing human contact networks with
different contact characteristics. To illustrate the impact of
disruptions in the IL on the evolution of an outbreak, we utilized two
different edge removal protocols: (i) random edge removal and (ii)
targeted edge removal. In both protocols, we select a proportion $q$
of nodes and then remove corresponding edges with probability $p$. In
random edge removal, we select nodes in the IL uniformly at random while we
select nodes with the largest degree (\ie, hub nodes) in targeted edge
removal. Although edge removal may render the IL disconnected, the EL is always connected in our simulations such that all nodes in the EL can potentially become infected. Previous work has shown that scale-free networks such as the
IL in our multiplex network are more robust to random than targeted
disruptions.\cite{cohen2000resilience, cohen2001breakdown,
  albert2000error} The reason for this effect is that by removing hub
nodes of a scale-free network, a large number of all edges in the
network is being removed, strongly impacting the connectivity
properties of such a network. We observe that targeted edge removal
can abruptly change outbreak characteristics such as time to peak
infection, even for small proportions of selected nodes. Our results
extend those presented in previous work on random and targeted
disruptions\cite{cohen2000resilience, cohen2001breakdown,
  albert2000error} by establishing a connection to coevolving
information and epidemic diffusion.
\section*{Data availability}
Our source codes are publicly available at \url{https://gitlab.com/ComputationalScience/information-epidemic}.
\bibliography{aipsamp}
	
\end{document}